\documentclass[a4paper]{article} 
\usepackage{epsfig}
\usepackage{amsmath}
\usepackage[left=2cm,top=2cm,bottom=3cm]{geometry}

\makeatletter 
\@addtoreset{equation}{section} 
\makeatother 
 
\newcommand{\be}{\begin{equation}} 
\newcommand{\ee}{\end{equation}} 
\newcommand{\bea}{\begin{eqnarray}}
 \newcommand{\eea}{\end{eqnarray}}
 \def\non{\nonumber }
\newcommand{\ov } {\over } 
\newcommand{\p }{\partial }
 \newcommand{\s }{\sigma }
 \def\half{{\texstyle {1\over 2}}}
 
\def\r{r }

\def\a{\alpha }

\def\w{\omega }
\def\lb{\lambda}
\def\lbb{\lambda^ 2}
 \def\vareps{\varepsilon }
\def\half{{\textstyle {1\over 2}}}

\def\appendix#1{   \addtocounter{section}{1}   \setcounter{equation}{0}   
\renewcommand{\thesection}{\Alph{section}}   \section*{Appendix \thesection\protect\indent \parbox[t]{11.15cm}   {#1} }   \addcontentsline{toc}{section}{Appendix \thesection\ \ \ #1}   } 

\def\appendix#1{
  \addtocounter{section}{1}
  \setcounter{equation}{0}
  \renewcommand{\thesection}{\Alph{section}}
  \section*{Appendix \thesection\protect\indent \parbox[t]{11.15cm}
  {#1} }
  \addcontentsline{toc}{section}{Appendix \thesection\ \ \ #1}
  }

\begin{document}

\null\vskip-24pt 
   \hfill UB-ECM-PF/07/30
      \vskip-1pt
\hfill {SISSA-76-2007-EP}
\vskip 1truecm
\begin{center}
\vskip 0.2truecm {\Large\bf New non-abelian effects on D branes 
 }
\vskip 0.2truecm

\vskip 0.7truecm
\vskip 0.7truecm

{\bf Roberto Iengo$^a$ and Jorge G. Russo$^{b,c}$}\\
\vskip 0.4truecm
\vskip 0.4truecm

${}^a${\it  International School for Advanced Studies (SISSA)\\
Via Beirut 2-4, I-34013 Trieste, Italy} \\
{\it  INFN, Sezione di Trieste}

\medskip

$^{b}${\it 
Instituci\' o Catalana de Recerca i Estudis Avan\c{c}ats (ICREA)

  \medskip

$^{c}${\it 
Departament ECM,
Facultat de F\'\i sica, Universitat de Barcelona,\\
Diagonal 647, 08028 Barcelona, Spain} 

} 
 
\end{center}
\vskip 0.2truecm 

\noindent\centerline{\bf Abstract}

We extend the Myers dielectric effect to configurations with angular momentum.
The resulting time-dependent $N$ D0 brane bound states can be interpreted as describing rotating fuzzy ellipsoids.
A similar solution exists also in the presence of a RR magnetic field, that we study in detail.
We show that, for any finite $N$, 
above a certain critical angular momentum is energetically more favorable for the 
bound state system to dissociate into an abelian configuration of $N$ D0 branes moving independently.
We have investigated this problem in the low-energy expansion of the non-abelian D brane action 
for generic $N$. 
In the case $N=2$ 
we  find  explicit solutions of the full non-abelian Born-Infeld D brane dynamics,
which remarkably have the same structure and confirm the features of the low-energy approximation. 
We further study D string configurations representing fuzzy funnels deformed by the magnetic field and by 
the rotational motion.


\newpage

\renewcommand{\thefootnote}{\arabic{footnote}}
\setcounter{footnote}{0}

\section{  Introduction}

\setcounter{equation}{0}

The study of the D brane dynamics led to many important results in the last years.
When $N$ D branes are put together, a nonabelian $U(N)$ gauge symmetry emerges \cite{witten}.
The scalar fields describing the transverse displacements of the branes become matrix-valued
in the adjoint representation of the gauge group. This leads to solutions describing 
non-commutative configurations where branes of lower dimensions can generate fuzzy higher dimensional branes.
The basic example is the Myers ``dielectric effect" \cite{myers}. In this case $N$ D0 branes
moving in an external four-form field reduce their energies by forming a bound state, whose geometry  can be
recognized as the geometry of a fuzzy sphere.
The system can be interpreted as a fuzzy version of the bound state between a spherical D2 brane and $N$ D0 branes, and a precise correspondence exists between the corresponding energies and radii in the large $N$ limit.

Another interesting example of  non-abelian effects in D brane systems are the ``funnel" solutions
found in \cite{constable1,constable2}, representing bound states of D1 and D3 branes.
The non-abelian solutions appear in the $N$ D1 brane system even in the absence of external fields.
Remarkably, the dynamics produces a fuzzy version of the BIon system,  spike solutions 
where D strings extend out of the D3 brane \cite{callan,howe,gibbons}.
A review of  different non-abelian phenomena can be found in \cite{myers2} and various interesting applications can be found e.g. in 
\cite{trivedi}--\cite{thomas2}.

Finding more general non-abelian solutions  is in general  complicated due to the non-linear
nature of the equations. 
In this paper we will find a class of time-dependent exact solutions, which is remarkably simple
despite the non-linear effects.
We will also turn on an external magnetic RR field, generalizing a study made in \cite{ILR} for a single
Dp brane moving in an external RR magnetic field to the case of $N$ branes.
We will first consider  the case of the $N$ D0 brane system, and then show that similar time-dependent
solutions exist for the funnel solutions of \cite{constable1,constable2}.

When the magnetic field is absent, our D0 brane system is equivalent to the one studied by Myers \cite{myers}, where
$N$ D0 branes move under the influence of an external RR four-form field strength $F^{(4)}_{t123}$.
In our case it will be convenient to take $F^{(4)}_{t123}$ and $F^{(4)}_{t456}$, in order to satisfy the Gauss constraint in the presence of angular momentum.
The static system of Myers  describes $N$ D0 branes forming a bound state representing a fuzzy 2-sphere geometry (for further discussions on these geometries see e.g. \cite{dewit,madore}).
Here we will have the same system to start with, in terms of $\Phi^i_+\equiv \Phi^i+\Phi^{i+3}$, $i=1,2,3$ 
(in the static limit our system is {\it not} the fuzzy version of $S^2\times S^2$ 
studied in \cite{trivedi}).
We will first study what happens when angular momentum is given to the system and then add the magnetic field.
The structure of the time-dependent solution is similar with or without the magnetic field, and the main 
features are as follows.
At sufficiently small angular momentum, the energy of the non-abelian system is lower than the energy of $N$ D0 branes moving independently under the action of
the magnetic field. The non-abelian system represents a fuzzy ellipsoid.
As the angular momentum is increased, there is a critical point where the energy of the non-abelian 
bound state overcomes the energy of the abelian configuration of $N$ independent D0 branes.
We interpret this as a signal that the $N$D0-D2 brane bound state must break for angular momenta above a certain critical value, that we calculate.

Here we will use the non-abelian D brane action derived in \cite{myers}. 
We will consider  the flat space background $G_{\mu\nu}=\eta_{\mu\nu},\ B_{\mu\nu}=0$, $\phi={\rm const.}$
and the world-volume gauge field strength $F_{\a\beta}=0$.
We will turn on RR gauge fields and neglect the back reaction on the metric.
This is justified for weak fields. More precisely, one assumes that the fields are sufficiently weak so that
the corresponding energy density multiplied by the Newton constant is much smaller than one.

{}For flat backgrounds, the action of \cite{myers} reduces to $S=S_{BI}+S_{WZ}$,  where
\bea
{S}_{BI}=-T_p \int d^{p+1}\sigma\, {\rm STr}\left( \sqrt{\det(Q^i{}_j)}
\sqrt{-\det\left(
P\left[G_{\alpha \beta}+G_{\alpha i}(Q^{-1}-\delta)^{ij}G_{j\beta}\right]\right)} \right), \label{BIuno} 
\eea 
$$
Q^i{}_j\equiv\delta^i{}_j+i\lambda\,[\Phi^i,\Phi^k]\,G_{kj}
$$
and the Wess-Zumino term is
\begin{equation}
S_{WZ}=T_p\int {\rm STr}\left(P
\left[e^{i\lambda\, i_\Phi i_\Phi} \sum C^{(n)}\right]\right) \ .
\label{BIcs}
\end{equation}
As usual, $\Phi^ i$ represents the transverse displacements, $\Delta X^i=\lb \Phi^i\ ,\ \lb=2\pi l_s^2$.
For further details we refer to \cite{myers,myers2}.


\section{D0 brane bound states }
\setcounter{equation}{0}

\subsection{Time-dependent non-abelian solutions}

The low-energy Lagrangian for $p=0$ in the presence of an external RR field strength $F^{(4)}$ is given by
\be
L= -NT_0+ T_0\lbb {\rm Tr}\left({1\over 2} \dot \Phi^2 +{1 \over4} 
[\Phi^i,\Phi^j]\,[\Phi^i,\Phi^j]
+{i\over3}\, \Phi^i\Phi^j\Phi^k
F^{(4)}_{tijk}(t)\right)+O(\lb^ 4)\ ,
\label{lagD0}
\ee
where the the non-zero components of the field strength are $F^{(4)}_{t123}= F^{(4)}_{t456}=-2f$, i.e.
\be
F^{(4)}_{tijk}=-2f \vareps_{ijk}\qquad{\rm for}\ \{ i,j,k\} =1,2,3\ \ {\rm or}\ \  \{ i,j,k\} =4,5,6\ .
\ee
The equations of motion are
\be
-\ddot \Phi^ i + [\Phi^j,[\Phi^ i,\Phi^ j]]-  i f \vareps_{ijk} [\Phi^ j,\Phi^ k]=0\ ,\qquad i,j,k=1,2,3\ ,
\label{eqmot}
\ee
and similarly for $i,j,k=4,5,6$. They have to be supplemented by the Gauss constraint
\be
\sum_{i=1}^9 [\dot \Phi^j,\Phi^j] =0\ ,
\label{gss}
\ee
coming from the equation of motion of the gauge field $A_0$, which was set to zero. Having a field strength
with non-vanishing components in both directions 123 and 456 will permit to solve the constraint due to a balance
between the plane 12 and the plane 45.
With this choice of field strength components, the Myers solution  \cite{myers} is given by 
\be
\Phi^k = \Phi^{k+3}={f\over 2}\ \a^k \ ,\ \qquad k=1,2,3\ ,
\ee
where $\alpha^k $ are two $N \times N$ matrix representations of
the SU(2) algebra
\be
[\a^i,\a^j]=2i\,\vareps_{ijk}\,\a^k\ ,
\ee
Introducing new variables $\Phi^i_\pm = \Phi^i\pm \Phi^{i+3}$, the $\Phi^i_+$ satisfy the commutation relations defining a fuzzy two-sphere:
\be
[\Phi_+^i,\Phi_+^j]=2i r_0\,\vareps_{ijk}\Phi_+^k\ ,
\qquad r_0={f} \ .
\label{fuzzy}
\ee
while other commutators vanish (he gauge field configuration also admits a solution representing a fuzzy $S^2\times S^2$ configuration, where the $\Phi^{456}$ are given in terms $\a^{456}$, commuting with the $\a^{123}$ \cite{trivedi}).

We now look for time-dependent solutions to this system.
We consider the following ansatz:
\bea
\Phi^1 &=& k_1(t)\, \a_1 +  k_2(t)\, \a_2\ , \qquad \Phi^4 = k_1(t)\, \a_1 -  k_2(t)\, \a_2\ , \non\\
\Phi^2 &=& q_1(t)\, \a_1 +  q_2(t)\, \a_2\ , \qquad \Phi^5 = -q_1(t)\, \a_1 +  q_2(t)\, \a_2\ , \non\\
\Phi^3 &=& \Phi^6= m_0\, \a_3\ .
\label{ggj}
\eea
The equations of motion reduce to
\bea
&& -\ddot k_1 -8 (k_2^2+q_2^2+ m_0^2) k_1 + 4 f m_0 q_2 =0\ ,\non\\
&& -\ddot k_2 -8 (k_1^2+q_1^2+ m_0^2) k_2 - 4 f m_0 q_1 =0\ ,\non\\
&& -\ddot q_1 -8 (k_2^2+q_2^2+ m_0^2) q_1 - 4 f m_0 k_2 =0\ ,\non\\
&& -\ddot q_2 -8 (k_1^2+q_1^2+ m_0^2) q_2 + 4 f m_0 k_1 =0\ ,\non\\
&& -2(k_1^2+k_2^2+q_1^2+q_2^2)m_0+ f(k_1q_2-k_2 q_1)=0\ .
\eea
Although the system is non-linear, it admits a very simple solution:
\bea
k_1 &=& r_0\ \cos(\w t)\ ,\qquad k_2=r_0\ \sin(\w t)\ ,\non\\
q_1 &=& - r_0\ \sin(\w t)\ ,\qquad q_2=r_0\ \cos(\w t)\ ,\non\\
m_0 &=& {f\over 4}\ ,
\label{hhj}
\eea
with $\w $ defined by the equation
\be
\w^ 2= 8r_0^ 2- {1\over 2}f^2 \ .
\label{aass}
\ee
The Gauss constraint (\ref{gss}) is identically satisfied. 

Introducing  $\Phi^ i_\pm = \Phi^i\pm \Phi^{i+3}$, we find 
\bea
\Phi^1_+ &=& 2 r_0\cos(t\w )\a_1\ ,\qquad \Phi^2_-=-2r_0\sin(t\w )\a_1\ ,\non\\
\Phi^2_+ &=& 2 r_0\cos(t\w )\a_2\ ,\qquad \Phi^1_-=2r_0\sin(t\w )\a_2\ ,\non\\
\Phi^3_+ &=& {f\over 2}\ \a_3\ ,\qquad \Phi^3_-=0\ ,
\eea
They describe the algebra of a rotating fuzzy ellipsoid. To visualize this, at any given time $t$
one can make a rotation of coordinates so that the algebra is the same as the one at $t=0$,
where $\Phi^-_{1,2,3}$ vanish and $\Phi^+_{1,2,3}$ 
 satisfy the following commutation relations:
\be
t=0:\ \ \ [\Phi^1_+,\Phi^2_+]=16i {r_0^2\over f}\, \Phi^3_+\ ,\qquad 
[\Phi_+^3,\Phi_+^1]=i f \, \Phi^2_+\ ,\qquad 
[\Phi_+^2,\Phi_+^3]=i f\, \Phi^1_+\ . 
\label{fuzell}
\ee
At generic $t$, the ellipsoid rotates penetrating into the $\Phi^{1,2}_-$ part of the space.
The eccentricity of the ellipsoid 
is always the same and determined by the ratio
$4r_0/f$. When this is equal to one, one gets from (\ref{aass}) $\w=0$ and we recover the Myers static solution.
From (\ref{aass}) we see that $4r_0/f$ must be greater than one in order to have oscillatory solutions.

In the large $N$ limit it should describe a bound state of $N$ D0 branes and a D2 brane.
The classical configuration can be visualized by replacing the $\a_1, \a_2, \a_3$ matrices 
with spherical harmonics $Y_{1,1},\ Y_{1,-1},\ Y_{1,0}$, i.e.
\bea
X^1 &=& k_1(t)\  \cos(\varphi)\sin(\theta)+k_2(t)\ \sin(\varphi)\sin(\theta)\ , \non\\
X^2 &=& q_1(t)\ \cos(\varphi)\sin(\theta)+q_2(t)\ \sin(\varphi)\sin(\theta)\ ,\non\\
X^3 &=& {f\over 4}\ \cos(\theta)\ ,
\eea
and a similar configuration in $X^4,X^5,X^6$. This surface should represent the dual D2 brane, though we were unable to prove it
from the dual D2 brane system, which for time-dependent configurations becomes complicated.

The energy of the solution is given by
\be
E=N T_0 +{1\over 32} \lbb T_0 c_N   \big( -f^4 + 4f^2\w^2 + 12\w^4 \big)  \ , 
\label{huno}
\ee
where we used
\be 
{\rm Tr}[\a^i \a^ j] =c_N \delta^ {ij}\ ,\ \qquad c_N={N\over 3}(N^2-1)\ ,\ \ i,j=1,2,3\ ,
\label{normali}
\ee
which holds for irreducible representations (for reducible representations one gets a coefficient less than $c_N$).
The formula (\ref{huno}) shows that the rotational motion increases the energy of the system, as expected.

\subsection{Adding a RR magnetic field}

Consider an external RR field $C^{(1)}$, with non-vanishing components
 $C_1^{(1)}=\half b\lb \Phi^2 ,\ C_2^{(1)}=-\half b\lb \Phi^1$ and 
similar ones in the plane 4,5, i.e. $C_4^{(1)}=-\half b\lb \Phi^5 ,\ C_5^{(1)}=\half b\lb \Phi^4$.
This can be produced by  D6 brane and  anti D6 brane configurations.
{}From eq. (\ref{BIcs}), one can read the additional term in the lagrangian
\be
\Delta L = T_0 \lb {\rm Tr}\big[ C_i^{(1)}\dot \Phi^ i\big]=   T_0 \lbb  \half 
b {\rm Tr}[ \Phi^2\dot \Phi^1- \Phi^1\dot \Phi^2 - \Phi^5\dot \Phi^4 + \Phi^4\dot \Phi^5]\ .
\label{mgmg}
\ee
The equations of motion now become
\bea
&&-\ddot \Phi^i + [\Phi^j,[\Phi^ i,\Phi^ j]]-  i f \vareps_{ijk} [\Phi^ j,\Phi^ k]- \vareps_{ij3} b \dot \Phi^ j=0\ ,
\ \ \ i=1,2\ ,\non \\
&&-\ddot \Phi^ 3 + [\Phi^j,[\Phi^ 3,\Phi^ j]]-  i f \vareps_{3jk} [\Phi^ j,\Phi^ k] =0\ .\ 
\label{ebbb}
\eea
The equations for $i=4,5,6$ are similar with $b\to -b$.
The constraint is
\be
\sum_{i=1}^9 [P^i,\Phi^i]=0\ ,\
\label{gerbo}
\ee
where $P^i =\dot \Phi^i +\half \epsilon^{ij} b\Phi^j $ for $i,j=1,2$ and $P^{\bar i} =\dot \Phi^{\bar i} -\half \epsilon^{{\bar i}{\bar j}} b\Phi^{\bar j} $,  $\bar i,\bar j=4,5$ and otherwise $P^i=\dot \Phi^i$.
The simplest solution is obtained by considering all $\Phi^i$ to be diagonal, with entries $(x^ i_1,...,x^i_N)$,
representing the positions of $N$ D0 branes. In this case they move independently in the magnetic field,
giving rise to the usual Landau motion of the form 
$x_\a^ 1= \rho_\a \cos b t, \ x^2_\a=\rho_\a \sin b t$, $x_\a^ 4= \rho_\a \cos b t, \ x^5_\a=-\rho_\a \sin b t$,
$\a=1,...,N$.

Here we will be interested in non-abelian solutions. Once we turn on non-diagonal components of the $\Phi^ i$,
then only the center-of-mass coordinate --proportional to the $N\times N$ identity-- follows the Landau motion,
where the $N$ D0 branes move collectively under the magnetic force.
This motion decouples, since the identity commutes with all other matrices of $U(N)$, so in what follows
we consider the $SU(N)$ part only, i.e. traceless $\Phi^ i$.

We consider the same ansatz (\ref{ggj}), which now leads to the equations
\bea
&& -\ddot k_1 -8 (k_2^2+q_2^2+ m_0^2) k_1 + 4 f m_0 q_2 -b \dot q_1=0\ ,\non\\
&& -\ddot k_2 -8 (k_1^2+q_1^2+ m_0^2) k_2 - 4 f m_0 q_1 -b \dot q_2=0\ ,\non\\
&& -\ddot q_1 -8 (k_2^2+q_2^2+ m_0^2) q_1 - 4 f m_0 k_2+ b\dot k_1  =0\ ,\non\\
&& -\ddot q_2 -8 (k_1^2+q_1^2+ m_0^2) q_2 + 4 f m_0 k_1+ b\dot k_2  =0\ ,\non\\
&& -2(k_1^2+k_2^2+q_1^2+q_2^2)m_0+ f(k_1q_2-k_2 q_1)=0\ .
\eea
A solution is given again by (\ref{hhj}), where now
\be
\w^ 2+b\w- 8r_0^ 2+{1\over 2} f^2=0 \ ,
\label{bwww}
\ee
i.e.
\be
\w=-{1\ov 2}\big( b \pm \sqrt{b^ 2 +32r_0^2-2 f^ 2}\ \big) \ .
\label{bzzz}
\ee
The constraint (\ref{gerbo}) is identically satisfied.

The commutation relations of the $\Phi^ k,\ k=1,...,6$ are the same as in (\ref{fuzell}).
We see that a sufficiently large magnetic field permits to have oscillatory solutions for arbitrarily small values of the eccentricity $4r_0/f$.

Another interesting feature is that there are two solutions that represent a spherical configuration $r_0={f\over 4}$.
One is the obvious solution $\w =0$. This is the Myers solution, which survives even in the presence of the magnetic field, since a static configuration does not feel the magnetic force. A second solution is $\w =-b $. This is an oscillatory solution which remarkably remains spherical.

The energy of the general solution (\ref{hhj}), (\ref{bzzz}) is given by
\be
E=N T_0 + \lbb T_0 c_N \big(2r_0^2\w^2+8r_0^4
-r_0^2f^2 \big)\ ,
\label{nbel}
\ee
where we used (\ref{normali}).

Let us compare the energy of  the  spherical configuration $r_0={f \over 4}$ for the two cases: i) $\w= 0$ and  ii) $\w=-b $. In these cases the energy formula (\ref{nbel}) gives
\be
E_{\rm (i)}=N T_0 -{1\over 32}  \lbb T_0 c_N f^4 \ ,\qquad E_{\rm (ii)}=E_{\rm (i)}+ {1\over 8}  \lbb T_0 c_N b^ 2 f^ 2  \  .
\label{iiii}
\ee
The first case with energy $E_{\rm (i)}$ is the same as the result of \cite{myers}.  We see that the second case has more energy.
More generally,
using that $r_0^2>0$ and $\w^ 2>0$ one can see that the solution with less energy is the Myers static solution
 $r_0={f \over 4}$  and energy $E_{\rm (i)}$. To see this, one can rewrite the energy (\ref{nbel}) in the form
\be
E=N T_0 +{1\over 8}\lbb T_0 c_N \big(- {1\over 4} f^ 4 +\w^ 2  (b+\w)^ 2+ 16r_0^ 2 \w^ 2  \big)\ ,
\label{nbb}
\ee
 where we used the relation (\ref{bwww}).

The energy can also be expressed in terms of the conserved angular momenta defined by
\be
J_{12}=T_0\lbb {\rm Tr} \big[\dot \Phi_1 \Phi_2- \dot \Phi_2 \Phi_1+\half b(\Phi_1^2+\Phi_2^2)\big] \ ,
\ee
\be
J_{45}=T_0\lbb {\rm Tr} \big[\dot \Phi_4 \Phi_5- \dot \Phi_5 \Phi_4-\half b(\Phi_4^2+\Phi_5^2)\big] \ .
\ee
The non-abelian properties of the dynamics provides
a contribution to an ``intrinsic spin" of the system, to be distinguished from the contribution of the center-of-mass, which
represents the orbital part.

It is interesting to compare the energy of the abelian solution
where the $N$ D0 branes move independently under the action of the magnetic field with
the present non-abelian solution, for the same value of the angular momentum.
As mentioned above, the abelian solution is given by 
$(\Phi_1,\Phi_2,\Phi_4,\Phi_5)={\rm diag}(\rho_1,...,\rho_N)\times (\cos(b\tau ),\sin(b\tau ),\cos(b\tau ),-\sin(b\tau ))$.
The angular momentum of the abelian solution is given by
\be
J_{12}^{\rm abel} =-J_{45}^{\rm abel} = -\half N T_0\lbb \rho_0^ 2 b\ ,\qquad \rho_0^2\equiv{1\over N}(\rho_1^2+...+\rho_N^2)\ ,
\ee
and the energy has the expected gyromagnetic contribution
\be
E_{\rm abel}= N T_0 +  N T_0\lbb \rho_0^ 2 b^ 2=N T_0 +|b J_{12}|+|b J_{45}|\ .
\label{ebel}
\ee

In the case of the non-abelian solution, the angular momentum is given by
\be
J_{12}^{\rm nonabel}=-J_{45}^{\rm nonabel}= T_0\lbb c_N \ j \ ,\qquad j= r_0^ 2 (2\w +b)\ .
\label{jna}
\ee
To simplify the discussion, we assume that the center of mass does not move, i.e. that in the non-abelian case
all the angular momentum comes
from the intrinsic spin.

Using (\ref{bzzz}) and (\ref{jna}) we find the relation
\be
{j}=\pm r_0^2\sqrt{b^ 2 +32r_0^2-2 f^ 2}\ .
\label{jr0}
\ee
The energy can then be written in the form
\be
E_{\rm nonabel}= NT_0 + \lbb T_0 c_N \big( -8 r_0^4 + {j^2\over r_0^2} - {b j}\big)\ .
\label{kkjj}
\ee
The first term  $NT_0$ represents as usual the contribution from the rest masses of the $N$ D0 branes.
The second term proportional to $- 8 r_0^4 $ is the binding energy. The
next term is a kinetic contribution due to the rotation and finally the term $- \lbb T_0 c_N {b j}= - {1\over 2}b (J_{12}-J_{45})$
corresponds, interestingly, to a gyromagnetic interaction with gyromagnetic factor equal to 1 (so that  
the magnetic moments of the non-abelian system are equal to ${1\over 2}J_{12},\ {1\over 2}J_{45}$).  


The energy (\ref{kkjj}) of the non-abelian solution has a complicated expression in terms of the angular momentum. The reason is that
$r_0^2$  also depends on the angular momentum through (\ref{jr0}) and it 
is determined  by a cubic equation.
Since $|j|$ increases monotonically with $r_0^2$, there is a one-to-one correspondence
between $|j|$ and $r_0^2$ and therefore the value of $j$ univocally determines the value of the energy.

{}Using (\ref{jr0}) one finds  that at large $|j|$ the energy has the expansion:
\be
E_{\rm nonabel}= NT_0 + \lbb T_0 c_N \big( {3\over 2^{1/3}} \, |j|^{4/3} -{1\over  2^{5/3}} \, |j|^{2/3}(f^2-{b^2\over 2}) - {b j} -{1\over 96}(f^2-{b^2\over 2})^2+...\big)
\label{yyt}
\ee
In the special case $f^2= b^2/2$ we get $r_0^2=2^{-5/3}|j|^{2/3}$ and $\w =-{b\over 2}+{\rm sign}(j)\, (4|j|)^{1/3}$ and
one can write a simple expression for the complete energy
\be
E_{\rm nonabel}\bigg|_{f^2= {b^2\over 2}}
=NT_0 +  c_N T_0\lbb  \big(- b j+ 3\cdot 2^{-1/3} |j|^{4/3}\big)\ .
\ee

Since the energy (\ref{yyt}) for the
non-abelian solution increases as $|J|^{4/3}$ for large $|J|$ (recall $J= T_0\lbb c_N j$),  there must be a point where
it overcomes the linearly-increasing energy (\ref{ebel}) of the abelian configuration.
This indicates the remarkable fact that there is a critical value of the angular momentum after which 
the Myers D2-$N$D0 bound state breaks and the system prefers to
behave as $N$ individual D0 branes.

The critical $J$ can be found as follows. 
Writing $J_{\rm abel}=J_{\rm nonabel}\equiv T_0\lbb c_n j$,
the abelian energy can be written as
\be
E_{\rm abel}=N T_0 +2T_0\lbb c_N |b j|\ .
\ee
Expressing $j$ in terms of $r_0$ as in (\ref{jr0}), we see from (\ref{kkjj})  that 
$E_{\rm abel}=E_{\rm nonabel}$
for
\bea
&& 24r_0^2-2 f^2+{b^2} = b\sqrt{32r_0^2-2{f^2}+{b^2} }\ ,\ \qquad j<0
\non\\
&& 24r_0^2-2 f^2+{b^2} =3 b\sqrt{32r_0^2-2{f^2}+{b^2} }\ ,\ \qquad j>0
\eea
This gives a unique solution for $r_0^2$. Substituting into eq. (\ref{jr0}) we find the critical angular momentum after which
the bound state should become unstable:
\bea
j_{\rm cr}&=&
\frac{1}{216}\left( -b^2 + 6\,f^2 + b\,{\sqrt{b^2 + 6\,f^2}} \right) \,{\sqrt{5\,b^2 + 6\,f^2 + 4\,b\,{\sqrt{b^2 + 6\,f^2}}}}
\ ,\qquad
j<0 \non\\
j_{\rm cr}&=& \frac{1} {24{\sqrt{3}}}    \left( 5\,b^2 + 2\,f^2 + b{\sqrt{33\,b^2 + 6\,f^2}} \right) {\sqrt{23 b^2 + 2f^2 + 4b{\sqrt{33\,b^2 + 6\,f^2}}}}
 \ ,\ 
j>0. 
\eea
Note that the $J_{\rm cr}= T_0\lbb c_N j_{\rm cr}$ grows like $N^3$ for large $N$.
This suggests that in the dual D2 system  
this transition may not be seen, since the critical point moves
to infinity as $N\to\infty $.

We stress  that this transition exists also in the Myers system with $b=0$, i.e. without the RR magnetic field. 
In this case 
$j_{\rm cr}\to \pm  {1\over 6\sqrt{6}} |f|^3$
 and the energy at the critical point goes to zero.
{}For $b=0$, the abelian configuration describes $N$ D0 branes with uniform motion.
After the transition, the $N$ D0 branes
should carry the energy and angular momentum of the original system in terms of linear momenta 
$\vec p_\a $, $\a=1,...,N$.

Let us now examine the behavior of the energy as $j\to 0$. From eq.(\ref{jr0}) we see that
$j=0$ corresponds to $r_0^2=(2 f^2 -b^2)/32$ if $f^2\geq b^2/2$ or $r_0=0$ if $f^2< b^2/2$.
Therefore, for $j=0$ the energy (\ref{kkjj}) takes the following form
\bea
E_{\rm nonabel}& \to & NT_0 -{1\over 32} c_N T_0 \lbb \Big( f^2- {b^2\ov 2}\Big)^2\ ,\ \ \ \ f^2\geq {b^2\over 2}\ ,
\non\\
E_{\rm nonabel}& \to & NT_0 \ ,\ \ \ \ f^2< {b^2\over 2}\ .
\eea
In Figures 1a, 1b, 1c we compare the  energies for each solution as a function of the angular momentum, including the abelian case. The figures exhibit the point $j_{\rm cr}$ of the transition after which 
$E_{\rm abelian}<E_{\rm nonabelian}$.

\begin{figure}[ht]\label{fig.1a,b,c}
\ \ \epsfig{file=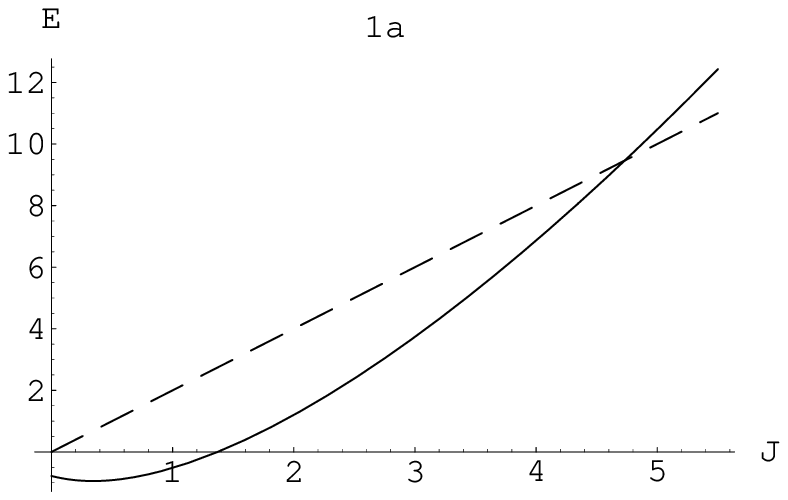,width=.4\textwidth}  \\
\medskip
\epsfig{file=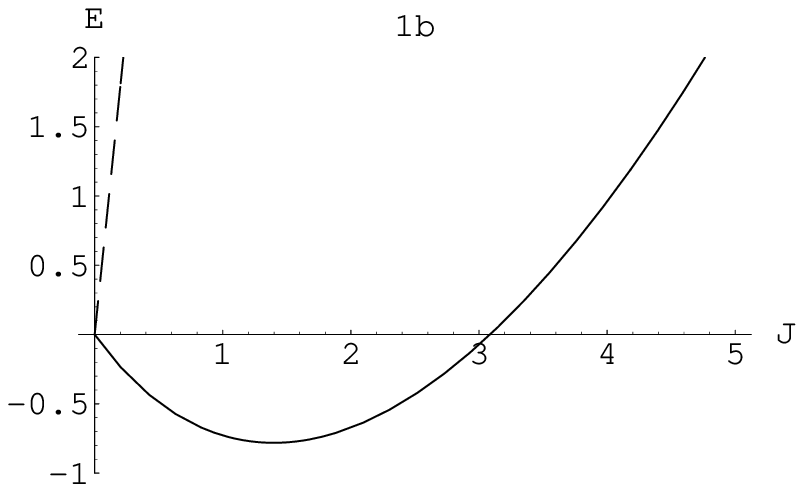,width=.4\textwidth}
\ \qquad \epsfig{file=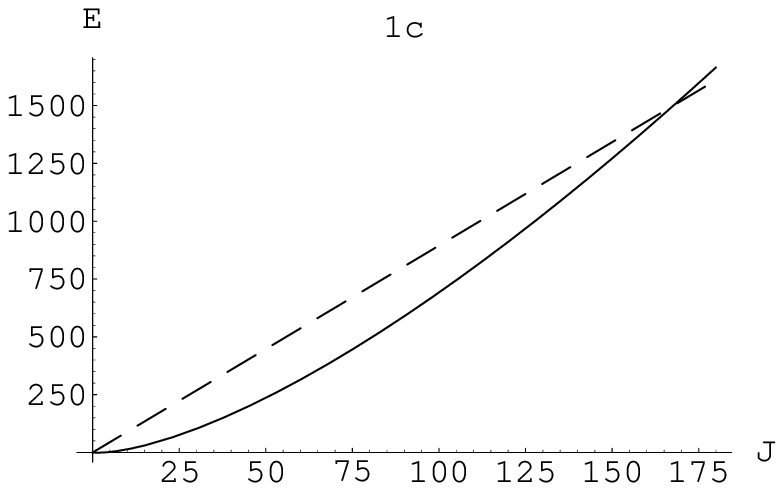,width=.4\textwidth}
\caption{Energy (with the constant part $NT_0$ substracted) for the nonabelian (solid line) and abelian (dashed line) solutions as  a function of the angular momentum for
the cases a) $f^2>b^2/2$; b) and c)  $f^2<b^2/2$. In figure a) the energy 
is negative  at $J=0$, where the solution
reduces to the Myers solution. Then it first decreases and then increases until it overcomes the energy of the abelian solution.
In the case b) the energy 
is zero at $J=0$; then it first decreases becoming negative and then
increases until it overcomes the abelian energy, at much larger values of $J $, as shown in figure 1c.}
\end{figure}

\smallskip
 
Finally, it is interesting to note that the
spherical case $r_0=f/4$, $\w= -b$, with energy $E_{\rm (ii)}$ given in (\ref{iiii}), 
 can be attained for a particular $j=-bf^2/16$.

\subsection{Corrections from Born-Infeld }

We will now show that the same ansatz promotes to a solution of the full Born-Infeld Lagragian
with only a correction in the equation defining the frequency and the constant value of $\Phi_3,\ \Phi_6$,
parametrized by $m_0$. Here for simplicity we will consider the case of $SU(2)$, so $N=2=c_N$ in what follows.
The non-abelian Born-Infeld action is a non-linear functional of the non-abelian fields
and a complete definition requires a prescription for the ordering of the fields under the trace.
The action (\ref{BIuno}) was defined with the symmetric trace prescription \cite{tseytlin}, that we also adopt.
For practical matters, this prescription amounts to treat $\dot\Phi^i$ and $i\big[ \Phi^i,\Phi^j\big]$
as commuting objects in the entries of the determinant.
{}For the D0 brane system, the BI part of the Lagrangian can be written as
\be
L_{\rm BI}= -T_0 {\rm Tr}\sqrt{-\det \tilde D}\ ,\qquad 
\tilde D =
\left(\begin{matrix}
- {\bf 1}_{2\times 2} & \lb \dot \Phi^j \\ -\lb \dot \Phi^i  & \delta_{ij}{\bf 1}_{2\times 2}+i\lb [\Phi^i,\Phi^j]
\end{matrix}\right)
\label{llll}
\ee
We now consider the following ansatz:
\bea
\Phi^1 &=& r_0\cos(\w t )\, \a_1 +  r_0\sin(\w t)\, \a_2\ , \qquad \Phi^4 =\bar r_0\cos(\bar\w t )\, \a_1 +  
\bar r_0\sin(\bar\w t) \ , \non\\
\Phi^2 &=& -r_0\sin(\w t)\, \a_1 +  r_0\cos(\w t)\, \a_2\ , \qquad \Phi^5 =-\bar r_0\sin(\bar \w t)\, \a_1 +  \bar r_0\cos(\bar \w t) \ , \non\\
\Phi^3 &=& \bar m_0\, \a_3\ ,\qquad\qquad \Phi^6= \bar m_0\, \a_3\ .
\label{ggbbb}
\eea
We will compute the Lagrangian on this ansatz and find the values of $\w ,\ m_0$ and 
$\bar r_0,\ \bar \w ,\ \bar m_0$ by variational principle with respect to $\r_0,\ m_0,\ \bar r_0,\ \bar m_0$
and by imposing the Gauss constraint, which in these variables reads
\be
\left({\p\over \p\w}+ {\p\over \p\bar \w}\right) L_{tot}=0
\ee
where $L_{tot}$ includes the same couplings to $F^{(4)}$ and $C^{(1)}$ given in sects. 2.1 and 2.2. 
We have performed several checks that this procedure reproduces the correct equations.

Invariance under rotation of the Lagrangian independently in the planes 12 and 45 implies
that we can compute $L_{tot}$ at $t=0$.
For $SU(2)$ 
we find that the $2\times 2$ matrix $\det \tilde D$ is proportional to the identity, so taking the square root
is straightforward (in the case of $SU(N)$ there would be terms proportional to $(\a_i)^2$ inside the square root which are not necessarily
proportional to the identity).
The complete Lagrangian is given by
\be
L_{tot}= -2T_0 \sqrt{-\det\tilde D}
+
2 \lb^2 T_0\left( (4 f  m_0 +b \w) r_0 ^2+ \bar r_0  ^2 (4 f  \bar m_0  - b \bar \w  )\right)\ ,
   \ee
   \bea
-\det \tilde D &=& 1+\lb^2\left(4 r_0 ^4+8 \bar r_0  ^2 r_0 ^2-2 \w^2 r_0 ^2+4 \bar r_0  ^4-2 \bar r_0  ^2 \bar \w  ^2+8
   m_0 ^2 \left(r_0 ^2+\bar r_0  ^2\right)+8 \bar m_0  ^2 \left(r_0 ^2+\bar r_0  ^2\right)\right) 
\non\\
&-&8\lb^4
   r_0 ^2 \bar r_0  ^2 \left(2 m_0 ^2+2 \bar m_0  ^2+r_0 ^2+\bar r_0  ^2\right) (\w-\bar \w  )^2
 \eea

The equations of motion and the constraint are solved by
\be
\bar r_0=r_0\ ,\qquad \bar m_0=m_0\ ,\qquad \bar \w=-\w\ .
\ee
In addition, they lead to the following equations:
\be
0=f-\frac{4 m_0 \sqrt{1 - 4\lb^2r_0^2\w^2}}{\sqrt {1+16 \lbb r_0^2\left(2 m_0^2+ r_0^2\right) } } \ ,
\label{unozz}
\ee
\be
0=4 b m_0\w (1 - 4\lb^2 r_0^2 \w^2) + f(8m_0^2 - 8r_0^2 + \w^2 + 48\lb^2 r_0^4\w^2)\ . 
\label{dosz}
\ee

Solving (\ref{unozz}) for $m_0$ we find 
\be
m_0={f\over 4}\ \frac{ \sqrt{1+  16 \lbb r_0^4}}{ \sqrt{1-2\lbb f^2 r_0^2-4 \lbb \w^2 \r_0^2}}\ .
\label{mzzz}
\ee
Combining (\ref{unozz}) and (\ref{dosz}) one can get an algebraic equation  determining 
the frequencies  $\w $.
One can then substitute $\w$ in eq. (\ref{mzzz}) and find  $m_0$.
Expanding in powers of $ \lbb $ we recover the same solution of section 2.1, as expected:
\be
 m_0 = \frac{f}{4}+O(\lb^2)\ ,\qquad \w^ 2+b\w- 8r_0^ 2+{1\over 2} f^2=O(\lb^2)\ ,
\ee
where we used (\ref{mzzz}) and (\ref{dosz}).

\subsubsection{Angular momentum and energy}

The angular momenta are:
\bea
J_{12} &=& 
{\p L_{tot}\ov\p \w }=2 T_0 \lambda^2 r_0^2 (b+ {8\omega m_0\ov f})\ .
\non\\
J_{45}&=& {\p L_{tot}\ov\p \bar \w }=-2 T_0 \lambda^2 r_0^2 (b+ {8\omega m_0\ov f})=-J_{12}\ .
\eea
The energy is:
\bea
E &=&  \w {\p L_{tot}\ov\p  \w}+\bar \w {\p L_{tot}\ov\p  \bar \w} -L_{tot}
\non\\
&=& T_0 { f(1 + 16\lb^2r_0^4(1 + 8\lb^2 m_0^2\w^2))\over 2m_0(1 - 4\lb^2r_0^2\w^2)}\ .
\eea
where we used (\ref{unozz}). 
Expanding in powers of $\lb $, to order $\lbb $ we reproduce the previous energy (\ref{nbel}) with $c_N=N=2$.

\subsubsection{System without magnetic field}

In the particular case $b=0$, the equations simplify. We get
\be
m_0= {f\over 4} \  \sqrt{1+48  \lb ^2 r_0 ^4}
\ ,\ \qquad
\w^2 = \frac{8 r_0 ^2-{1\over 2} f ^2 -24 \lb ^2 f ^2  r_0 ^4}{1+48 \lb ^2 r_0 ^4} \ .
\ee
and the angular momentum and the energy are given by
\bea
J&\equiv &J_{12}= -J_{45}=2T_0\sqrt{2}\lb^2 r_0^2\sqrt{16r_0^2 - f^2(1 + 48\lb^2 r_0^4)}
\ ,
\nonumber \\ 
E&=& 2T_0(1 - 2f^2\lb^2 r_0^2)\sqrt{1 + 48\lb^2 r_0^4}\ .
\label{erat}
\eea
The energy (\ref{erat}) reproduces, as expected, the formula (\ref{huno}), with $\w $ given by (\ref{aass}), 
modulo terms $O(\lb^4)$.
The possible values of $r_0$ and $f$ are constrained by the condition that $J$ and $E$ must be real.
For $\lb f^2>2/\sqrt{3}$ the quantity under the square root, $16 r_0^2-f^2-48\lb^2 f^2 r_0^4$
is negative and therefore $J$ becomes complex, so in what follows we assume $\lb f^2\leq 2/\sqrt{3}$. 
The minimum and the maximum values of $r_0$ are  given by the value at which $J=0$, i.e. defined by the condition $16 r_0^2-f^2-48\lb^2 f^2 r_0^4=0$. 
One can see that $J$ has a maximum value; for larger values
of the angular momentum, the solution does not exist and the bound state must break.

For $J$ less than the maximum value,  $E-2T_0$ can be negative (corresponding to a bound state) or positive.
This can be seen more clearly by expressing $J$ and  $E-2T_0$ as functions of the scaling variable 
$x\equiv r_0^2/f^2$ 
in the allowed interval $x_m\leq x\leq x_M$, at fixed values
of the parameter $p\equiv \lb^2 f^4 < 4/3$. Except when $p$ is near to the endpoint $p=4/3$, 
 for small $x-x_m$ (implying small $J$) one sees that $E-2T_0<0$ and it becomes positive for larger $x$ 
(larger $J$), implying that there should be a transition to a configuration of $N$ independent (abelian) D0 branes. 
Curiously, for larger $x$,  $E-2T_0$ remains positive even though $J\to 0$ for $x\to x_M$.
This $J=0$ configuration with $x=x_M$ and $\w=0$ represents an unstable equilibrium radius
of the Myers fuzzy 2-sphere, which was unnoticed in \cite{myers}.  
  For $p$ near the end point at $p= 4/3$, $J$ is small in the whole interval in $x$, where $E-2T_0<0$ for every $x$
so the bound state is stable in this small $J$ interval. For larger $J$ the bound state must break, as pointed out above.

This confirm the instability 
of the Myers bound state at sufficiently high angular momentum, including the full BI dynamics.



\section{ D-strings and funnels in a magnetic field}

Consider the Born-Infeld action (\ref{BIuno}) in the case $p=1$,
for flat space-time without any external field.
In the static gauge $t=\tau $, $\s=x^9$, to 
the leading order in the expansion in powers of $\lb $ one gets the equation
\be
\partial^ a\partial_a \Phi^i = [\Phi^j,[\Phi^j,\Phi^i ]]
\ee
In \cite{constable1,constable2} a non-trivial solution representing the dual of the BIon solutions \cite{callan,howe,gibbons} of the D3 brane
system was found. The ansatz is
\be
\Phi^i =\hat R(\s )\ \a^i\ ,
\ee
giving the equation
\be
\hat R'' = 8 \hat R^ 3\ ,
\label{eqfu}
\ee
with the solution 
\be
\hat R = \pm {1\over 2(\s -\s_\infty )}\ .
\label{solf}
\ee
This solution has a funnel shape where the cross section is a fuzzy sphere,
which is consistent with the interpretation as a dual of the BIon solution.
The fuzzy sphere has physical radius
\be
R(\s )= \lb \left( {1\over N} \sum_{i=1}^3 {\rm Tr}[\Phi^i(\s )^2]\right)^{1/2}=
{N\pi l_s^2\over (\s-\s_\infty )}\ \sqrt{1-{1\over N^2}}
\ .
\ee

\medskip
Consider now an external RR magnetic field $C^{(2)}$, with non-vanishing components
 $C_{1,9}^{(2)}=\half b\lb \Phi^2 ,\ C_{2,9}^{(2)}=-\half b\lb \Phi^1$,
$C_{4,9}^{(2)}=-\half b\lb \Phi^5 ,\ C_{5,9}^{(2)}=\half b\lb \Phi^4$ (recalling $x^9=\s $),
producing the  additional term in the lagrangian
\be
\Delta L = T_1 \lb \int d\s \ {\rm Tr}\big[ C_{i9}^{(2)}\dot \Phi^ i\big] =   T_1 \lbb  \half 
b \int d\s \ {\rm Tr}[ \Phi^2\dot \Phi^1- \Phi^1\dot \Phi^2-\Phi^5\dot \Phi^4 + \Phi^4\dot \Phi^5]\ .
\ee
We have set the gauge field compontents $A_0=A_1=0$, which implies the constraints
\be
\sum_{i=1}^9 [\partial_\sigma \Phi^i,\Phi^i]=0\ ,\qquad \sum_{i=1}^9 [P^i,\Phi^i]=0\ ,
\label{derf}
\ee
where $P^i$ is the momentum conjugate to $\Phi^i$ which has the same expression as in sect. 2.2.
We consider the following ansatz:
\bea
\Phi^1 &=& h_1(\s ) k_1(t)\, \a_1 + h_2(\s ) k_2(t)\, \a_2 \ , \qquad 
\Phi^4 =h_1(\s ) k_1(t)\, \a_1 - h_2(\s ) k_2(t)\, \a_2 \ ,\non\\
\Phi^2 &=& l_1(\s) q_1(t)\, \a_1 +  l_2 (\s)q_2(t)\, \a_2\ , \qquad
\Phi^5 = -l_1(\s) q_1(t)\, \a_1 +  l_2 (\s)q_2(t)\, \a_2\ ,\non\\
\Phi^3 &=& \Phi^6= m(\s) \, \a_3\ ,
\label{ggjz}
\eea
where $k_1, k_2, q_1, q_2 $ are given by eq. (\ref{hhj}) (setting $r_0=1$ which is superfluous in the presence of the arbitrary functions $h_i,l_i$).
Then it is easy to show that the constraints (\ref{derf}) are identically satisfied and that the
equations of motion imply $h_1 =l_1$ and $h_2=l_2$.
The equations of motion then give
\bea
&&h_1'' -8 h_1 (h_2^2+m^2)+\w(\w+b) h_1=0\ ,\non\\
&&h_2'' -8 h_2 (h_1^2+m^2)+\w(\w+b) h_2=0\ ,\non\\
&&m''-8 m (h_1^2+h_2^2)=0\ .
\label{rew}
\eea
In the particular case  $\w=0$ and $m=h_1=h_2\to \hat R$ we recover the solution (\ref{solf}) found in \cite{constable1,constable2}.
The same solution exists also in the time-dependent case when $\w=-b$.

Equations (\ref{rew}) follow from the effective Lagrangian:
\be
L_{\rm eff} = {h_1'}^2+{h_2'}^2+{m'}^2 +8 (h_1^2h_2^2+h_1^2 m^2+h_2^2m^2)-\w (\w+b) (h_1^2+h_2^2)\ .
\ee
The conservation of the ``energy" gives the first order condition
\be
E_0={h_1'}^2+{h_2'}^2+{m'}^2-8 (h_1^2h_2^2+h_1^2 m^2+h_2^2m^2)+\w (\w+b) (h_1^2+h_2^2)\ .
\ee
The solution with $b=0$ and $\w =0$ given in \cite{constable1,constable2} has $E_0=0$.
We will now consider a time-dependent solution parametrized by $\mu \equiv  \w (\w+b)$, 
with $h_1=h_2=h$ and with $E_0=0$. 
In the region where $h(\s ), \ m(\s )$ are large, the term multiplied by $\mu $ can be neglected as
compared to the non-linear term in eq. (\ref{rew}), and one finds a  solution of the form (\ref{solf}) 
with $h\cong m\cong \hat R(\s )$, for small $\s-\s_{\infty}$ 
(adopting similar initial conditions for $h$ and $m$). 
As $\s $ increases, the term proportional to $\mu $ becomes important producing a splitting between $h$ and $m$.
The equations can be solved numerically. The results can be summarized as follows.

\begin{itemize}

\item For $\mu >0$ the ratio $h/m$ --representing the eccentricity of the fuzzy ellipsoid-- decreases until a minimum value at $\s=\s_1$, giving a rugby-ball shaped ellipsoid (see fig. 2a).
This point represents the endpoint of the string, since for $\s>\s_1$  the solution becomes complex. 

\item For $\mu <0$ the  ratio $h/m$  increases until it diverges at $\s=\s_2$. At this point $m\to 0$ and
the solution degenerates into pizza-shaped ellipsoid (see fig. 2b). For $\s>\s_2 $  the solution becomes complex. 

\end{itemize}

\begin{figure}[ht]\label{fig.2a,b}
\centerline{(2a)\epsfig{file=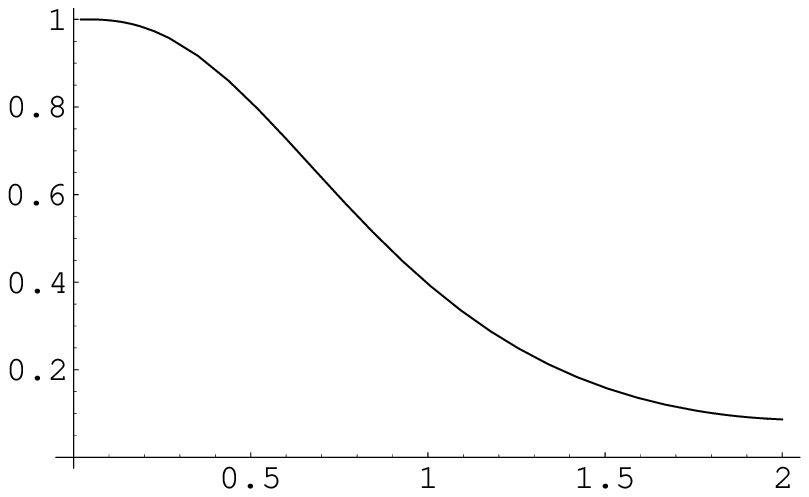,width=.4\textwidth}\ \ \qquad (2b)\epsfig{file=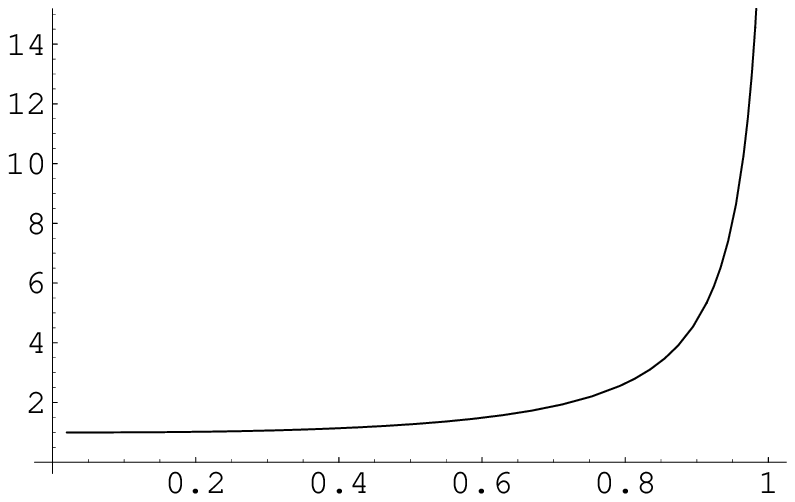,width=.4\textwidth}}
\caption{Plot of $h/m$ as a function of $\sigma$ for: a) $\mu=0.5$ and b) $\mu=-0.5$.}
\end{figure}


In summary, the rotational motion converts the fuzzy sphere into an ellipsoid with an eccentricity
$h/m$ that can be less or greater than one, according to the sign of $\mu =\w(\w+b)$. 
The eccentricity is different at each point of the string, becoming equal to one at $\s =0$. In the dual picture, 
this is the point where the $N$ D strings intersect the D3 brane.

\section*{Acknowledgments}

This work is also supported by the European
EC-RTN network MRTN-CT-2004-005104. J.R. also acknowledges support by MCYT FPA
2004-04582-C02-01 and CIRIT GC 2005SGR-00564. 

\vfill\eject\null

\end{document}